\documentclass [12pt]{article}

\usepackage {graphicx}

\begin{document}

\bigskip

\textbf{FLUCTUATION EFFECTS IN MIXED STATE OF TYPE II SUPERCONDUCTOR.}  

\textit{Comparison of Experimental Results with Theoretical Predictions}.

\bigskip

A.V.NIKULOV

Institute of Microelectronics Technology and High Purity Materials, Russian Academy of Sciences, 142432 Chernogolovka, Moscow District, Russia.

\bigskip

\textbf{1. Introduction}

\bigskip

Investigation of the fluctuation phenomena changes habitual notion about the mixed state of type II superconductors. Before the fluctuation investigation one thought that the second-order phase transition from the normal state into the Abrikosov state takes place in the second critical field, $H_{{\rm c}{\rm 2}{\rm} }$ [1]. Now it is clear that a phase transition in $H_{{\rm c}{\rm 2}}$ is absent and the transition in the Abrikosov state, called now often as "vortex lattice melting", occurs below $H_{{\rm c}{\rm 2}{\rm }}$ [2]. Now it is widely recognized that the normal state with superconducting fluctuation, called as "vortex liquid", exists not only above but also below $H_{{\rm c}{\rm 2}}$ [2]. This becomes widely known in the last time, after the investigation of the fluctuation phenomena in high-T$_{{\rm c}{\rm}}$ superconductors (HTSC). Therefore most of scientists connect these changes of the habitual notion about the mixed state with the HTSC investigation first of all. Moreover some authors suppose that these changes of the notion relate to HTSC only because they think that the fluctuation effects in conventional superconductors are very small [3]. But it is not right. Moreover just the investigation of the fluctuation phenomena of conventional superconductors have shown first that the transition into the Abrikosov state occurs below $H_{{\rm c}{\rm 2}}$. It was made before the HTSC discovery. But these result are not enough known. Therefore I will write about these results first of all in this report. I will show that the results obtained on high quality samples of $YBa_{{\rm 2}}Cu_{{\rm 3}}O_{{\rm 7}{\rm -} {\rm x}{\rm} }$ repeat in the main our results obtained by the investigation of bulk conventional superconductors. In the end I will discuss the main unsolved problem  the nature of the transition into the Abrikosov state. 

\bigskip

\textbf{2. Fluctuation effects in bulk superconductors}

\bigskip

I will write about results of investigation which have been obtained in 1980 - 1984 years. Partly they have been published in that time [4-7], partly they have been published recently [8], and some results [9] have not been published for the present. The transition called now "vortex lattice melting" was observed in 1981 [5]. On figure 1 the result obtained in this paper on conventional superconductors $V_{{\rm 3}}Ge$ is compared with results obtained on high quality $YBa_{{\rm 2}}Cu_{{\rm 3}}O_{{\rm 7}{\rm -} {\rm x}}$ samples in many papers in last time [10-12]. In both case a change of the same resistive properties is interpreted as a transition. But in our work [5] this change was interpreted as a transition from the normal state with superconducting fluctuation (which we called as "one-dimensional state") into the Abrikosov state whereas in $YBa_{{\rm 2}}Cu_{{\rm 3}}O_{{\rm 7}{\rm -} {\rm x}{\rm} }$ this change is interpreted as vortex lattice melting [10-12]. Although the denomination "vortex lattice melting" has become very popular I think that it is bad denomination for this transition.

\begin{figure}

\includegraphics[width=5.63in,height=2.52in]{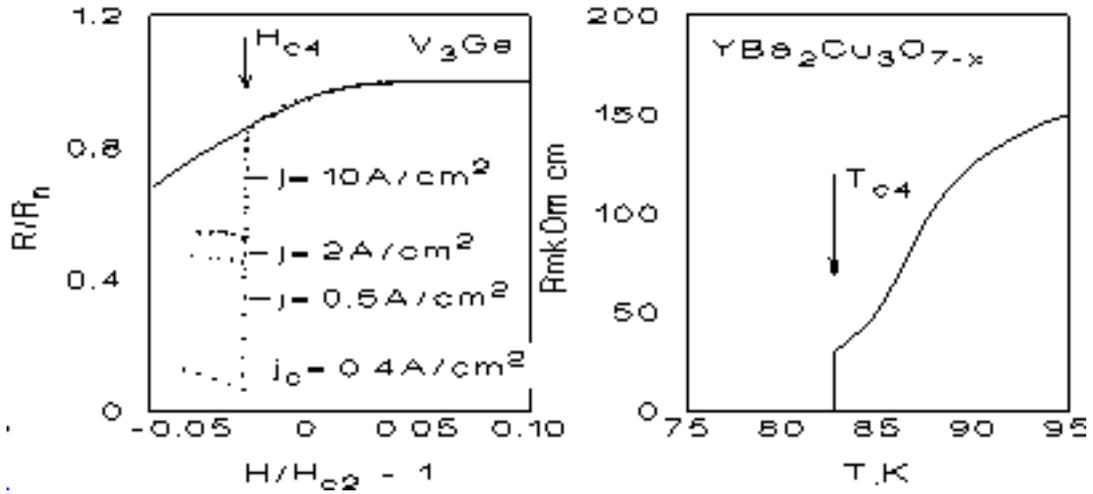}
 \caption{\textit{Resistive transitions in a perpendicular magnetic field of single crystals $V_{3} Ge$ [5] (points are experimental dependencies obtained at different current values, line is a theoretical paraconductivity dependence obtained in Hartree approximation) and $YBa_{2}Cu_{3}O_{7 - x}$ [10-12]}}
\end{figure}

The denomination "vortex lattice melting" assumes a existence of a other transition from the vortex liquid into the normal state. Many scientists think that the vortex liquid is the Abrikosov state also and a transition from the vortex liquid into the normal state exist. For example, in some works [13] the experimental dependencies of resistivity above the "vortex lattice melting" are compared with the flux flow resistivity dependencies. It can not be right. There does not exist a experimental base to think that two different transition take place on the way from the normal state into the Abrikosov state. Moreover our experimental results show that the sole transition from the normal state into the Abrikosov state is observed in conventional superconductors. 

The denomination "vortex lattice melting" has appeared from "a approach from bottom". In this approach the fluctuation in the Abrikosov state is considered. The Abrikosov state is considered as a vortex lattice. The fluctuations cause oscillations of the vortex. When the oscillations become enough big the vortex lattice melting takes place [14]. This image is very clear but this approach is quite no rigorous. The theoretical description of the Abrikosov state of the real superconductors is very difficult because the pinning effect plays very important role there. Therefore all theories of the vortex lattice melting base on different models which have many assumptions. In most of theories a superconductor without pinning is considered. And this is a main assumption that the Abrikosov state exists in superconductors without pinning. It is obvious from experiment [15] that the Abrikosov state exists in real superconductors but it has not been proved that the Abrikosov state exists in ideal superconductor without pinning centers. Moreover some authors are proving that the Abrikosov state can not exist in ideal superconductors [16-18]. If they are right the theories of vortex lattice melting are not right in base.

I will follow in my paper to "a approach from top". In the beginning I will write about the linear approximation region above $H_{{\rm c}{\rm 2}}$. After that I will come down in the critical region at $H_{{\rm c}{\rm 2}}$. Then I will tell about the changes which are observed at the transition from the normal state into the Abrikosov state. In the end of this part I will write about experimental results which we have obtained in the Abrikosov state. Because it is obvious now that the transition into the Abrikosov state takes place below $H_{{\rm c}{\rm 2}}$ we ought denote the position of this transition. In work [19] I have denoted the position of this transition as $H_{{\rm c}{\rm 4}}$. Below I will use by this denomination.

\bigskip

2.1. LINEAR APPROXIMATION REGION ABOVE $H_{{\rm c}{\rm 2}}$

\bigskip

Theoretical description of fluctuation in normal state is more easy than in the Abrikosov state. Therefore more rigorous theories were made for this region. Especially rigorous theoretical description was made for the linear approximation region [20]. This description is so reliable that on base of it we discovered first that the transition into the Abrikosov state takes place below $H_{{\rm c}{\rm 2} }$ [4]. Later we have confirmed this by a determination of $H_{{\rm c}{\rm 2}}$ from the magnetization measurement [7].

Our [4,6,8,9] investigation have shown that the experimental dependencies in linear approximation region are described very well by Ami-Maki theory [20]. The paraconductivity value along and across the magnetic field direction is anisotropic near $H_{{\rm c}{\rm 2} }$ [8,9]. The anisotropy decreases with removal from $H_{{\rm c}{\rm 2}}$. I want draw your attention the rigorous theory by Ami and Maki [20] only are valid for HTSC and "dirty" conventional superconductors [9]. The approximation theoretical dependencies used in most of works, for example obtained from popular Hikami-Larkin [21] and Lawrence-Doniac [22] works, do not have a validity region in HTSC. 

\bigskip

2.2. CRITICAL REGION

\bigskip

Our investigations have shown that in the critical region, the experimental dependencies of the magnetization and paraconductivity of bulk conventional superconductors can be plot as universal function of value $(h-h_{{\rm c}{\rm 2}})(ht)^{{\rm -2/3}}Gi^{{\rm -1/3}}$ and are close to a theoretical dependencies for a one-dimensional superconductor [6-9]. $t = T/T_{{\rm c}}$, $h = H/H_{{\rm c2}}(0)$, $H_{{\rm c2}}(0) = -T_{{\rm c}}(dH_{{\rm c2}}/dT)_{{\rm T = T_{c}}}$, $Gi = (k_{{\rm B}}T_{{\rm c}}/H_{{\rm c}}^{{\rm 2}}(0)\xi ^{{\rm 3}}(0))^{{\rm 2}}$ is the Ginzburg number of a bulk superconductor. This result repeated the result of the specific heat investigation near $H_{{\rm c2}}$ obtained in works [23,24]. These results have easy explanation if we use the lowest Landau level (LLL) approximation of the fluctuation Ginzburg-Landau theory. The thermodynamic average of any quantity, for example order parameter 

\begin{equation}
\label{eq1}
 < \vert \Psi \vert ^{2} > = {\frac{{\sum {\vert \Psi \vert ^{2}\exp ( - F_{GL} / k_{B} T)}} }{{\sum {\exp ( - F_{GL} / k_{B} T)}} }} 
\end{equation}

\noindent
is function of temperature, magnetic field value and superconductor parameters through the relation $F_{{\rm GL}}/k_{{\rm B}}T$ only. In a dimensionless unit system 

\begin{equation}
\label{eq2}
{\frac{{F_{GL}} }{{k_{B} T}}} = {\sum\limits_{k} {(\varepsilon _{n} + q^{2})\vert \Psi \vert ^{2} + {\frac{{1}}{{2V}}}{\sum\limits_{k_{i}}  {V_{k_{1} k_{2} k_{3} k_{4}}  \Psi _{k_{1}} ^{\ast}  \Psi _{k_{2}} ^{\ast}  \Psi _{k_{3}}  \Psi _{k_{4}} } } }} 
\end{equation}

\noindent
$\epsilon _{{\rm n}} = (t-1+h+2nh)/Gi_{{\rm H}}$; $Gi_{{\rm H}} = Gi^{{\rm 1/3}}(th)^{{\rm 2/3}}$ is Ginzburg number in magnetic field; $k = (n,q,l)$; $n$ is the number of the Landau level. 

According to (\ref{eq1}) and (\ref{eq2}) thermodynamic averages depend on temperatures, magnetic field value and superconductor parameters through $\epsilon  = \epsilon _{{\rm n = 0}}= (t-1+h)/Gi_{{\rm H}}$ only if terms with $n=0$ (LLL) take into account only in (\ref{eq2}). Thus, in the LLL approximation region 
the scaling law is general consequence of the fluctuation Ginzburg-Landau theory. 

The relation (\ref{eq2}) in LLL approximation is close to one for a one-dimensional superconductor. We may write this relation in following form

\begin{equation}
\label{eq3}
{\frac{{F_{GL}} }{{k_{B} T}}} = V[\varepsilon \overline {\vert \Psi \vert 
^{2}} + {\frac{{1}}{{V}}}{\int\limits_{V} {dr^{3}\Psi ^{\ast }{\frac{{d^{2}}}{{dz^{2}}}}\Psi} }  + {\frac{{\beta _{a}} }{{2}}}\overline {\vert \Psi \vert ^{2}} ^{2}] \end{equation}
 $\overline {\vert \Psi \vert ^{2}} $ is a spatial average value of order parameter. This expression is similar to one for one-dimensional superconductor. A main difference consists in a difference of a generalized Abrikosov parameter value $\beta _{a} = \overline {\vert \Psi \vert ^{4}} / \overline {\vert \Psi \vert ^{2}} ^{2}$ which is determined by a distribution of the order parameter in the space. Much below the transition $\Psi (r) = const$ in one-dimensional superconductor and consequently $\beta _{{\rm a}} = 1$. But function $\Psi (r) = const$ is not belong to LLL. (This has a simple interpretation: a magnetic field must penetrate through superconductor.) Therefore $min \beta _{{\rm a}} > 1$ for type II superconductor in a magnetic field. As it is well known Abrikosov has been found, in his famous work [25], $\beta _{{\rm a}}= 1.18$ for a square symmetry structure but later Kleiner, Roth and Autler [26] have shown that a triangular symmetry structure has a lesser value $\beta _{{\rm a}} \approx  1.16$. Now there are no reasons to doubt that $min \beta_{{\rm a}} \approx  1.16$ for the LLL approximation.

The difference $\beta _{a}$ value of the LLL and the one-dimensional superconductor is small ($\beta _{{\rm a}}$ changes from 2 above $H_{{\rm c2}}$ to $\approx 1.16$ below $H_{{\rm c2}}$ in the first case and from 2 above $T_{{\rm c}}$ to 1 below $T_{{\rm c}}$ in the second case). Therefore the dependencies of the quantities connected with a value of the spatial average order parameter (see (\ref{eq3})) (magnetization, specific heat and others) of bulk superconductors near $H_{{\rm c2}}$ are close to the one of a one-dimensional superconductor.

The LLL approximation is valid at $h \gg  Gi$; $h-h_{c2} \ll  2h$ and $h > h_{c2}/3$. For conventional superconductors a LLL crossover field $H_{{\rm LLL}} = H_{{\rm c2}}(0) Gi \approx 1 \ Oe$. It is much smaller than the experimental values of the magnetic field $10^{{\rm 4}} \ Oe$ used usually for mixed state investigation. But for HTSC the $H_{{\rm LLL}}$ value is much more: for $YBa_{{\rm 2}}Cu_{{\rm 3}}O_{{\rm 7 -x}}$, $H_{{\rm LLL}} \approx  10^{{\rm 4}} \ Oe$, for $BiSrCaCuO$, $H_{{\rm LLL}} \approx  10^{{\rm 5}} \ Oe$. Therefore the validity of the LLL approximation for HTSC is no so obvious as for conventional superconductors. However it was shown in some works [27] that the experimental dependencies of the magnetization and the specific heat of HTSC follow the scaling law obtained in the LLL approximation.

2.3. TRANSITION INTO THE ABRIKOSOV STATE

\bigskip

It is followed from the scaling law that in LLL region a transition must takes place at the same $\epsilon _{{\rm c4}}$ value for all superconductors. Our experimental investigations have confirmed this. We have measured [7] the $(H_{{\rm c4}} /H_{{\rm c2}} - 1)$ value in $V_{{\rm 3}}Ge$ and $Nb_{{\rm 94.3}}Mo_{{\rm 5.7}}$ with different value of the Ginzburg number. The relation $Gi_{{\rm VGe}}/Gi_{{\rm NbMo}}$ is equal approximately 1000 but for both superconductors the $\epsilon _{{\rm c4}} \approx  -1$ [19]. The $\epsilon _{{\rm c4}}$ value of $YBa_{{\rm 2}}Cu_{{\rm 3}}O_{{\rm 7-x}}$ is equal approximately -1 also. Thus now there are not reasons to think that the fluctuation effects in $YBa_{{\rm 2}}Cu_{{\rm 3}}O_{{\rm 7 - x}}$ differ heavily from the one of conventional bulk superconductors.

The transition in the $H_{{\rm c4}}$ connects with changes of resistive properties first of all. There are observed the following changes: 1) Above $H_{{\rm c4}}$ the current-voltage characteristics are Ohmic and below $H_{{\rm c4}}$ they are non-Ohmic. 2) Above $H_{{\rm c4}}$ the resistive dependencies are described by the paraconductivity theory and below $H_{{\rm c4}}$ they can not be described by this theory. 3) The main change: the resistive properties of enough homogeneous samples with different amount of pinning centers are universal above $H_{{\rm c4}}$ and they are non-universal below $H_{{\rm c4}}$. It is obvious that this changes connect with the vortex pinning appearance. 

From these results which we have obtained fifteen years ago [5] was clear that the transition into the Abrikosov state occurs below $H_{{\rm c2}}$. Our investigations of high homogeneous samples have shown that this transition much narrower than the "transition" on the specific heat, magnetization and on other properties connected with the spatial average order parameter. Thus, the experimental data indicate that two "transition" are observed [19]. It is obvious that the wide "transition" is a crossover connected with change of the spatial average order parameter and the narrow transition is connected with a change of a distribution of the order parameter in space (see values $\overline {\vert \Psi \vert ^{2}}$ and $\beta _{{\rm a}} $ in the relation (\ref{eq3})). The famous Abrikosov work [25] investigated the distribution of the order parameter in space (the value of the spatial average order parameter can be easy calculated (see relation (\ref{eq3})). Therefore the narrow transition (but no the wide crossover) ought be considered as the transition into the Abrikosov state. 

Above three changes are observed in $YBa_{{\rm 2}}Cu_{{\rm 3}}O_{{\rm 7 - x}}$ single-crystals also. We observed a fourth change also: 4) The flux flow resistivity decreases at $H_{{\rm c4}}$ [4,9]. Such changes of the flux flow resistivity we observed on all enough homogeneous superconductors which we measured. It is follow from these results that the theoretical dependencies for the flux flow resistivity obtained in the mean field approximation (see [28]) do not have a validity region in "dirty" superconductors.

Our measurement of resistivity along magnetic field and the measurement the non-local resistivity [8.9] have shown that any transition connected with break of symmetry along magnetic field [2] is not observed above $H_{{\rm c4}}$ in conventional superconductors. Thus, our results prove that the sole transition from the normal state into the Abrikosov state is observed near $H_{{\rm c2}}$ in conventional superconductors.

\bigskip

2.4. PROPERTIES OF THE ABRIKOSOV STATE OF REAL SUPERCONDUCTORS

\bigskip

Below $H_{{\rm c4}}$ the resistive properties of different samples are non-universal. The current-voltage characteristics are non-Ohmic and have different form for different samples. At big current the current-voltage characteristics of samples with a small critical current can be described by a relation $E = \rho _{{\rm f}}(j-j_{{\rm cd}})$, where $\rho _{{\rm f}}$ is the flux flow resistivity and $j_{{\rm cd}}$ is the dynamic critical current. But even the flux flow resistivity dependence is no quite universal for different samples [9].

The critical current dependencies $j_{{\rm c}}(H)$ of different samples are very different. $j_{{\rm c}}(H)$ of some samples have a pick near $H_{{\rm c4}}$ and the one of other samples do not have this pick [9]. The height of the pick was different in different samples. The critical current value in the pick reached 50 of this value before the pick [9].

The flux creep was observed in some samples near $H_{{\rm c4}}$ and was not observed in other samples [9]. We observed the change a curvature on a $logI-logV$ plot which was interpreted in some recent papers [29] as a transition from a vortex-glass into the vortex liquid. But our results [9] are explained by classical Kim-Anderson flux creep theory.

Our investigation have shown that the irreversibility line can not be interpreted as a position of a transition, because the position of this line is no universal for different samples [9]. The irreversibility line of samples with weak pinning is situated well below the $H_{{\rm c4}}$ [9]. The same was shown recently for $YBa_{{\rm 2}}Cu_{{\rm 3}}O_{{\rm 7-x}}$ [30]. 

In consequence of the flux creep the resistivity at small current may be very small below $H_{{\rm c4}}$ [9]. It causes a artifact in the magnetization dependence if a sample moving is used at the measurement [7,9]. In resent works [30,31] a "discontinuity" in the magnetization at the transition into the Abrikosov state was observed in $YBa_{{\rm 2}}Cu_{{\rm 3}}O_{{\rm 7 -x}}$ sample with weak pinning. This discontinuity was interpreted as a consequence of a first order transition. But this discontinuity may be the artifact caused by sharp decreasing resistivity of sample below $H_{{\rm c4}}$ as well as in conventional superconductors.

You see now that the state below$H_{{\rm c4}}$ is much more difficult for theoretical description than the state above $H_{{\rm c4}}$. It is very difficult to described a state in which properties are not universal. Therefore the "approach from bottom" is bad.

\bigskip

\textbf{3. About nature of the transition into the Abrikosov state}

\bigskip

It is clear from the experimental results what occurs in $H_{{\rm c4}}$. Above $H_{{\rm c4}}$ we have the normal state with superconducting fluctuation. Below $H_{{\rm c4}}$ the voltage at small current is equal zero if the flux creep is absent. According to the Josephson relation $\hbar d\varphi /dt = 2eV$, $V = 0$ means that the phase difference $\varphi $ between points of a sample is a constant in time. This means that above $H_{{\rm c4}}$ we can not conduct a line through 
superconducting region only and below $H_{{\rm c4}}$ we can make it. Thus, it is obvious that the phase coherence appears below $H_{{\rm c4}}$. But it is not known now what type is this transition. This problem is connected with a problem of existence of the Abrikosov state in ideal superconductors without pinning.

Results of our recent investigation of fluctuation effects in thin films of conventional superconductors [32] are important for the understanding the nature of the transition into the Abrikosov state. We have found that the position of the change of the resistive properties which we interpret as the transition into the Abrikosov state depend on a amount of pinning centers in thin films. In thin films with small amount of pinning centers this change do not observed up to very low magnetic field. The resistivity dependencies are described by the paraconductivity dependencies well below $H_{{\rm c2}}$. 

We interpret it as a dependence of a position of the transition into the Abrikosov state, $H_{{\rm c4}}$, on a amount of pinning centers of two-dimensional superconductors and absence of the Abrikosov state in ideal superconductors. In my recently work [18]  I have propose a concept of the transition into the Abrikosov state in two-dimensional superconductor according to which the position of this transition depend on a sample size or a amount of pinning centers.

\bigskip

\textbf{4. References}

\bigskip

1. De Gennes, P.G. (1966) \textit{Superconductivity of Metals and Alloys}, Pergamon Press.

2. Blatter, G., Feigel'man, M.V., Geshkenbein, V.B., Larkin, A.I., and Vinokur V.M. 
(1994) Vortex in High-T$_{{\rm c}}$ Superconductors, \textit{Rev.Mod.Phys.} \textbf{66}, 1125-.1388.

3. Fisher, D.S., Fisher, M.P.A., and Huse, D.A. (1991) Thermal fluctuations, quenched disorder, phase transitions, and transport in type-II superconductors, \textit{Phys.Rev.B} \textbf{43}, 130-159

4. Marchenko, V.A. and Nikulov, A.V. (1981) Fluctuation conductivity in $V_{{\rm 3}}Ge$ near the second critical field, \textit{Zh.Eksp.Teor.Fiz.} \textbf{80}, 745-750 (\textit{Sov.Phys.-JETP} \textbf{53}, 377-381).

5. Marchenko, V.A. and Nikulov, A.V. (1981) Magnetic field dependence of the electrical conductivity in $V_{{\rm 3}}Ge$ in the vicinity of $H_{{\rm c2}}$, \textit{Pisma Zh.Eksp.Teor.Fiz.} \textbf{34}, 19-21 (\textit{JETP Lett.} \textbf{34}, 17-19). 

6. Marchenko, V.A. and Nikulov, A.V. (1983) Paraconductivity of $V_{{\rm 3}}Ge$ in magnetic field, \textit{Fiz.Nizk.Temp}. \textbf{9}, 816-821. 

7. Marchenko, V.A. and Nikulov, A.V. (1984) Magnetization of type-II superconductors near the upper critical field $H_{{\rm c2}}$, \textit{Zh.Eksp.Teor.Fiz.} \textbf{86}, 1395-1399 (\textit{Sov.Phys.-JETP} \textbf{59}, 815-818).

8. Marchenko, V.A. and Nikulov, A.V. (1993) Anisotropy of the resistive 
transition width of conventional type II superconductors at the second critical field $H_{{\rm c2}}$, \textit{Physica C} \textbf{210}, 466-472.

9. Nikulov, A.V. (1985) Fluctuation phenomena in bulk type-II superconductors near second critical field, Thesis (Inst. of Solid State Physics, USSR Academy of 
Sciences, Chernogolovka).

10. Safar, H., Gammel, P.L., Huse, D.A., Bishop, D.J., Rice, J.P., and Ginzberg, D.M. (1992) Experimental evidence for a first-order vortex-lattice-melting transition in untwinned single crystal $YBa_{{\rm 2}}Cu_{{\rm 3}}O_{{\rm 7}}$, \textit{Phys.Rev.Lett.} \textbf{69,} 824-827.

11. Kwok, W.K., Fleshler, S., Welp, U., Vinokur, V.M., Downey, J., Crabtree, G.W., and Miller, M.M. (1992) Vortex lattice melting in untwinned and twinned single crystals of $YBa_{{\rm 2}}Cu_{{\rm 3}}O_{{\rm 7 - x}}$, \textit{Phys.Rev.Lett.} \textbf{69}, 3370-3373.

12. Jiang, W., Yeh, N.-C., Reed, D.S., Kriplani, U., and Holtzberg, F. (1995) Possible origin of anisotropic resistive hysteresis in the vortex state of untwinned $YBa_{{\rm 2}}Cu_{{\rm 3}}O_{{\rm 7}}$ single crystals, \textit{Phys.Rev.Lett}. \textbf{74}, 1438-1441.

13. Berghuis, P. and Kes, P.H. (1993) Two-dimensional collective pinning and vortex- lattice melting in $\alpha -Nb_{{\rm 1- x}}Ge_{{\rm x}}$ films, \textit{Phys.Rev.B} \textbf{47}, 262-272.

14. Nelson, D.R. and Seung, H.S. (1989) Theory of melted flux liquids, \textit{Phys.Rev. B} \textbf{39}, 9153-9174.

15. Essmann, U. and Trauble, H. (1967) Direct observation of a triangular flux-line lattice in type II superconductors by a Bitter method, \textit{Phys.Lett. A} \textbf{24,} 526-532.

16. O'Neill, J.A. and Moore, M.A. (1993) Monte Carlo investigation of the properties of the vortex liquid in two-dimensional superconductors, \textit{Phys.Rev. B} \textbf{48}, 374-391.

17. Tesanovic, Z. (1994) Critical behavior in type-II superconductors, \textit{Physica C} \textbf{220}, 303-309.

18. Nikulov, A.V. (1995) Existence of the Abrikosov vortex state in 
two-dimensional type-II superconductors without pinning, \textit{Phys.Rev. B} \textbf{52}, 10429-10432.

19. Nikulov, A.V., (1990) On phase transition of type-II superconductors into the mixed state, \textit{Supercond.Sci.Technol.} \textbf{3}, 377-380.

20. Ami, S. and Maki, K. (1978) Fluctuation-induced electrical conductivity in dirty type-II superconductors, \textit{Phys.Rev. B} \textbf{16}, 4714-4724.

21. Hikami, S. and Larkin, A.I. (1988) Magnetoresistance of high temperature superconductors, \textit{Modern Physics Lett.} \textbf{2}, 693-698.

22. Lawrence, J. and Doniach, S. (1970) Proceedings of 12th International Conference on Low Temperature Physics, Academic Press of Japan 361-363.

23. Hassing, R.F., Hake, R.R., and Barnes, L.J. (1973) Magnetic-field-induced one dimensional behavior in the specific-heat transition in dirty bulk superconductors, \textit{Phys. Rev.Lett.} \textbf{30}, 6-9.

24. Farrant, S.P. and Gough C.E. (1975) Measurement of the fluctuation heat capacity of niobium in a magnetic field, \textit{Phys.Rev.Lett.} \textbf{34}, 943-946. 

25. Abrikosov, A.A. (1957) On the magnetic properties of superconductors of the second group, \textit{Zh.Eksp.Teor.Fiz.} \textbf{32}, 1442-1452 (\textit{Sov.Phys.-JETP} \textbf{5}, 1174-1184).

26. Kleiner, W.H., Roth, L.M., and Autler S.H. (1964) Bulk solution of Ginzburg-Landaw equations for type II superconductors: Upper critical field region, \textit{Phys.Rev. A} \textbf{133}, 1226-1227.

27. Jendupeux, O., Schilling, A., Ott, H.R., and van Otterlo, A. (1996) Scaling of the specific heat and magnetization of $YBa_{{\rm 2}}Cu_{{\rm 3}}O_{{\rm 7}}$ in magnetic fields up to 7 T, \textit{Phys.Rev.} $B$ \textbf{53}, 12475-12480.

28. Gor'kov, L.P. and Kopnin, N.B. (1975) Flux flow and electrical 
resistivity of type II superconductors in a magnetic field, \textit{Usp.Fiz.Nauk} \textbf{116}, 413-448 (\textit{Sov.Phys.Usp.} \textbf{18}, 496-531.)

29. Koch, R.H., Foglietti, V., Gallagher, W.J., Koren, G., Gupta,A., and Fisher, M.P.A. (1989) Experimental evidence for vortex-glass superconductivity in Y-Ba-Cu-O, Phys.Rev. Lett. 63, 1511-1514. 

30. Liang, R., Bonn, D.A., and Hardy, W.N. (1996) Discontinuity of reversible magnetization in untwinned YBCO single crystals at the first order vortex melting transition, \textit{Phys.Rev.Lett.} \textbf{76,} 835-838. 

31. Welp, U., Fendrich, J.A., Kwok, W.K., Crabtree, G.W., and Veal, B.W. (1996) Thermodynamic evidence for a flux line lattice melting transition in $YBa_{{\rm 2}}Cu_{{\rm 3}}O_{{\rm 7 - x}}$, \textit{Phys. Rev.Lett.} \textbf{76,} 4809-4812.

32. Nikulov, A.V., Remisov, D.Yu., and Oboznov, V.A. (1995) Absence of the transition into Abrikosov vortex state of two-dimensional type-II superconductor with weak pinning, Phys.Rev.Lett. 75, 2586-2589.

\end{document}